\documentclass[conference]{IEEEtran}
\usepackage{graphicx}
\begin{document}

\title{Lessons Drawn from Implementation of Online Tutoring System in Physics Courses}

\author{ Itzhak Goldman\\
Department of Basic Sciences\\
Afeka, Tel Aviv Academic College of Engineering\\
Tel Aviv, Israel\\
 email: goldman@afeka.ac.il}  

\maketitle

\begin{abstract}
The online tutoring system CAPA was implemented at Afeka College in the academic year 2000-2001 in calculus based physics courses. It was used also in the academic year 2001-2002 and was very successful in improving understanding and achievements of the students.
The paper describes the system features and the case study of  its implementation. Lessons drawn from use of the CAPA system are discussed. The system no doubt contributed significantly to increased understanding and higher achievements in the final exams. This was at a price that students felt that they had to devote more
time to study than traditionally required. The instructor has to invest more time than traditionally done in composing  new problems and managing the system.  
However, the increased interaction with the students makes the teaching process much more interesting and rewarding.
 \end{abstract}

\section{Introduction}
   
The CAPA system (Computerized Asynchronous Personalized Assignments)
has been developed, in the 90s,  at Michigan State University (MSU) $[1]$.
The developers were physicists that confronted difficulties with teaching physics to large undergraduate  classes.   
The difficulties were 
\begin{itemize}
\item
The "loss" of the individual student in such a large class 
\item
Lack of real-time feedback to the lecturer 
\item
The not simple task of ensuring a uniform level for the  recital classes given  by teaching assistants 
\end{itemize}
The motivation was to develop a, simple to use, computerized system  that will supply personalized exercises  and  provide immediate grading to the students.
It turned out that the system had many additional benefits, notably improved achievements in exams$[2-6]$. This effect was more pronounced for the weaker students. The system's relative simplicity and reported success made it a favorite of
many universities and colleges in the US and it was used also in other disciplines, e.g. $[6, ]$.

Following the positive reports about CAPA I implemented it  at Afeka College in the academic year  2000-2001.  It was used in calculus based physics classes for engineering students. An additional motivation was due to the lack, at the time, of teaching assistants for correcting homework assignments. Given the success of the system, I have used it also
in the academic year 2001-2002. In the following  year, it was decided that a uniform format for the physics courses must be employed by the different instructors. Not all instructors were willing
to invest the effort required in using the system and preferred traditional teaching. This
brought to end the use of CAPA, {\it in spite of its success}. 

\section{System Features}

 The system was implemented on a Linux machine (Pentium 2)
running RedHat 6.  Support from the technical stuff at Afeka was very important.
  A passing note - the server proved extremely stable- running uninterrupted for months despite heavy load occurring  near the due time of the homework.

The system is:

 \begin{itemize}
 \item
A framework capable of delivering personalized assignments 
\item 
Content independent - the content is supplied by the instructor 
\item
Web interfaced - each student had a personal user number, and for each exercise sheet there was a capa-id consisting of 4 digits. The students could retrieve their capa-ids from the system 

\begin{figure}[b]
\fbox{
\centerline{
 \includegraphics*[scale=0.35]{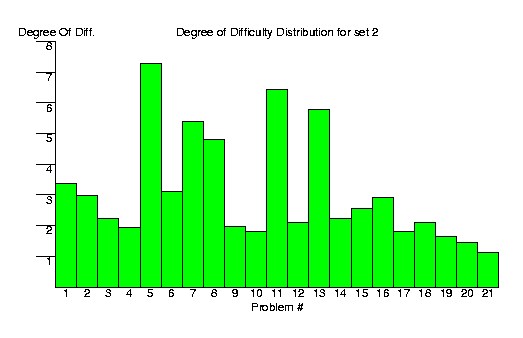} }}
\caption{Relative difficulty of problems.}
\end{figure}

\item
Asynchronous - allowing   students to pursue the assignment at their preferred time-schedule, subject to the constraint of due time 

\item
Immediate feedback is provided to the student on his performance 
\item
 A useful tool is a report assigning a measure of difficulty to each of the problems. This provides the lecturer with real-time feedback on what is well understood and what is not, so he can take corrective measures immediately. An example of such a report is shown in Fig. 1  
\end{itemize}

\begin{itemize}
 
\item
 Various statistical analyses, such as the grades distributions demonstrated in Fig. 2, can be readily obtained and provide real-time feedback 
\item
The system comes with a web forum that is useful in enhancing collaboration between the students and provides the instructor feed back on how well the students are doing 
\end{itemize}

\begin{figure}[h]
\fbox{
\centerline{
 \includegraphics*[scale=0.5]{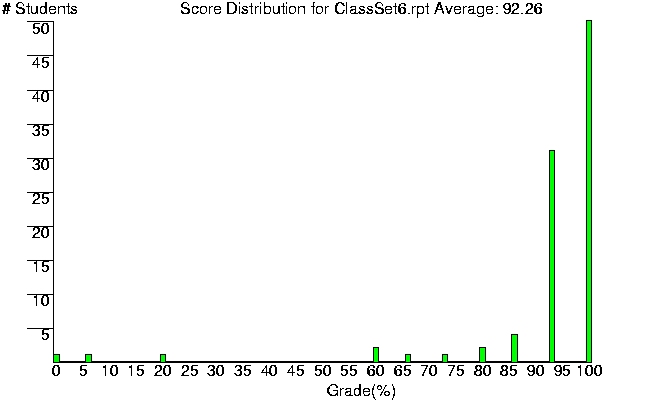} }}
\caption{Grades distribution for a given assignment.}
\end{figure}

\begin{figure} 
\fbox{
\centerline{
 \includegraphics*[scale=0.4]{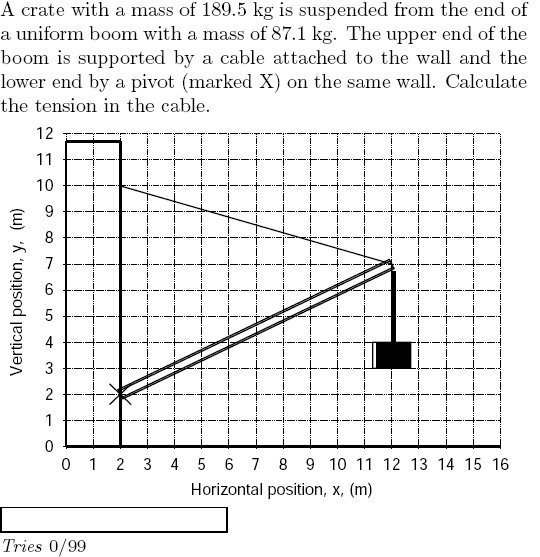} }}
 \caption{Numerical problem.}
\end{figure}

\begin{figure}[h]
\fbox{
\centerline{
 \includegraphics*[scale=0.5]{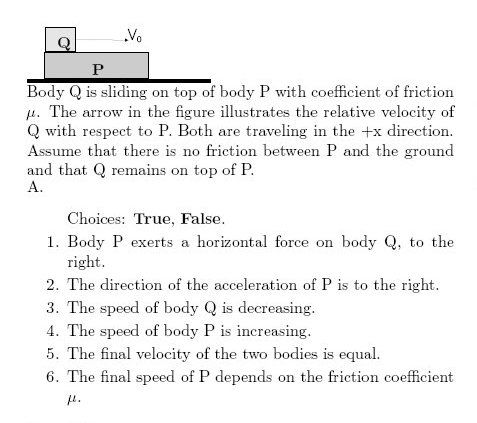} }}
\caption{Multiple choice qualitative problem.}
\end{figure}

\begin{figure}[h]
\fbox{
\centerline{
 \includegraphics*[scale=0.5]{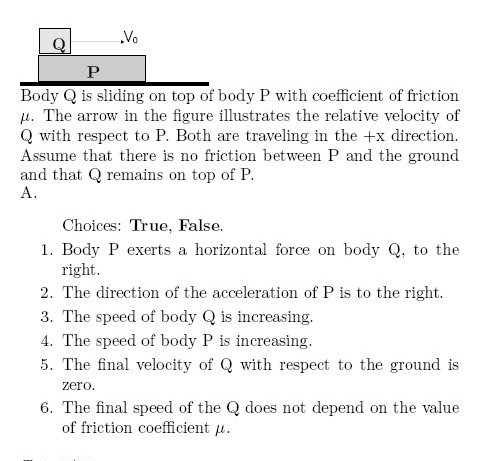} }}
\caption{The same problem for a different student.}
 
\end{figure}

 Each student receives the same problem but with {\it different} numerical data, or {\it different} questions to answer. These features promote cooperation and discussion among students while rendering mechanical copying much less likely
There are two main types of problems:
\begin{itemize}
\item
Numerical problems, as in Fig.3. Each student receives {\it different} numerical data 
  \item
Multiple-question qualitative problem that tests  understanding. Each student
 receives {\it different} combinations of the questions or altogether {\it different} questions relating to the same situation. Fig. 4 represents such a problem and Fig. 5 shows a variant
of it  

\item
The system responses
are Y, N or U. Y means a correct answer, N means an incorrect answer and U, (in case of the numerical problems) means wrong units. The allowed precision is set by the instructor. Typically, it was $1\div  2 \%$  
\item
Each problem can be tried several times. Normally, 10 tries were used. The multiple
tries facility was intended to compensate for misclicks or imprecise calculations.
The U replies do not count in the number of tries 
\item 
It is possible to build the problems so that after a few wrong answers, there will be a hint or a short explanation.  
 \end{itemize}

\section{  Case Study}
The system was implemented and used in the Physics 1 course; which is taken by students in their first semester of the academic studies. At 2000-2001, the physics lecturers were also giving the recital classes. There were two classes totaling of 86 students. 
There were 10 exercise sheets during a semester of 14 weeks. Overall, there were about 140 different problems, the best 120 answers were considered. Each problem had 10 tries and each problem set was open forf 7 -10 days depending on the work involved.
An online forum was set up. I reviewed it on a daily basis - reading comments, questions
and occasionally throwing in guiding clues.
 
\begin{itemize}

\item
At the beginning there was quite an enthusiasm from the students part.
They worked willingly and accordingly achieved very high scores on homework assignments.
I heard comments from lecturer's of other subjects that the students asked them why they are not employing CAPA too 
  
\item
At about mid-semester complaints had started. The fact that the system language is
English and not Hebrew was one of them  
\item
The fact that the system checks only final answers but not the detailed solution was regarded as unfair, even though the 10 tries availability should have compensated for calculation errors  
\item
However, the most serious complaint was that
they are forced to devote too much time to physics on the expense of other courses 
\end{itemize}
A survey that I conducted showed that they devoted $3\div 6$ hours weekly.
The course was 3 weekly hours of lectures and 2 weekly hours of recitals. In physics studies,  is usually
assumed that on the average a student should put 2 hours of self-study for each hour
of lectures. By this measure the time was more than reasonable, but the students truly felt that they are over-burdened. I'll discuss this point in the discussion section.

\begin{itemize}
\item
The good news were the success in the final exam: the grades were higher than in previous years - the median grade was higher by 8-10 points out of 100 - increasing  from $\sim 60$ to $\sim 70$ 
\item
Moreover, most of the improvement was  in the lower grades of the distribution.
This came as no surprise as similar results were reported in the US 
\end{itemize}

In the subsequent semester,   the same students took the course Physics 2. In response to the complaints in the previous semester, the students were offered two options: continue with CAPA as in the first semester, or have traditional
homework assignments. 
\begin{itemize}

\item
To my surprise (given the complaints) about  $80 \%$ chose CAPA 

\item
The students who chose CAPA said that in the final exam they felt that they were
much better prepared than in other courses and attributed it to CAPA. So in the new course there were willing to pay the price of investing extra time 

\item
In Physics 2, the grades in the final exam correlated positively with the use of CAPA. 
The small number of non-CAPA students renders statistical	 assessments not very significant; but the trend was qualitatively obvious 

 \item
Not few students continued to solve the problems even after they had 120 correct answers and they knew that the grade won't  increase.
A typical answer to the question why they did so was that they wanted to have a "perfect score" - from the system in addition to getting 100 from the instructor 
\item
Some students   regarded the CAPA assignments as fun - sort of a computer game- especially so because they received {\bf{\it immediate}} assessment 
\end{itemize}

As instructor, I benefited from the feedback utility that gave me in real-time a report on how the students   progressed  with each exercise sheet.
Moreover, I found out quickly which of the questions are in particular  difficult and could
explain once more the relevant material, before the exercise sheet was due.

The fact that each student received a personalized version of the sheet was very helpful with students that were absent from college because of justified reasons such as illness or reserve army duty.  To these students the sheets were opened and closed when they returned to the college.

Given the success of the system, I have used it also
in the academic year 2001-2002. In the following year, the physics courses  given by different instructors, had to be identical. Not all instructors were willing
to invest the effort required in using the system and preferred traditional teaching. This
brought to end the use of CAPA, {\it in spite of its success}.
\section{Discussion}

In what follows  aspects relevant to the student, to the instructor and to the academic institution will be addressed.
\subsection{The students}
 
\begin{itemize}
\item
Achievements as measured by scores of the final exam improved significantly  
\item Equally important was the fact that the students devoted more time for self study than they would have otherwise 
\item
Another advantage was that the students were encouraged to collaborate and discuss
the exercises, but mechanical copying was not simple as with traditional homework. 
\item
However, students felt (justifiably or not) that they are
required to work harder at the expense of other courses 
 \end{itemize}
The first point is impressive, especially so since not only were the scores higher but also the level of understanding  was higher.
Also  encouraging is the fact that weaker students seem to have benefited the most.
This last finding is in accord with similar findings in the US$[4, 5]$. In the US, it was also found $[5]$ that
the improvements among female students was larger than among male students. In our case the small number of female students didn't allow a meaningful test of this point.
\subsection{The instructor}
\begin{itemize}

\item
An important  advantage for the instructor is the ability to obtain  real-time feedback on the assimilation and understanding of the study matter by the students 

\item 
The instructor has to invest significantly more time than traditionally done  

\item
He should be willing to be involved in a computerized project where composing and checking out new problems is time consuming

\item
However, the increased interaction with the students makes the teaching process much more interesting and rewarding for the instructor compared with the standard way 
 \end{itemize} 
\subsection{The academic institution}
There is an obvious advantage in using the system for all classes, studying with different instructors the same course. Otherwise, some students will regard it as  unfair  that they are required to perform additional duties, while others
will consider unfair that they don't have access to the system.
\begin{itemize}
\item
The system can be quite valuable in many disciplines of engineering studies  
\item
 It is free and can be used under a GNU license 
\end{itemize}
 
A condition for  success is a {\it firm commitment} of the institution to provide the necessary support. This should translate to:
\begin{itemize}
\item
Establishing a support team that will handle technical problems and will adapt into the system, the academic material developed by the instructors  
\item
This will reduce the extra load from the instructors and will encourage even
computer-shy instructors to join 
 \item
A policy requiring all instructors, of the same course, to use the system 

\end{itemize}
\section{Concluding Remarks}
In Israel, students are typically older by 4-5 years than in the US or Europe. Most of
them work, at least part time, some are already married, and a some are parents.   

It is plausible that this is the reason for the complaints about the time burden.
In a traditional way of homework assignment they could devise all sort of strategies that will allow them not to put in the time required. With CAPA they were forced to invest the time.
 
This brings up the possibility
 that the improved achievements are not due directly to the use of technology
but indirectly through  its role as forcing the students to devote a minimal amount of time for self study.  

The system provides the student with immediate feedback, explains briefly the
relevant subject and gives an opportunity
to correct mistakes. This is a very positive educational process. 

A  related important element is that many students regarded working with the system as fun with elements of a game. For a generation brought up on computer games this may be indeed an advantage. 

In conclusion,   technology doesn't make the educational process easier, but can be very efficient in enhancing it.$[8]$.
\newpage
 \section {Acknowledgments}
I am indebted to the CAPA team at Michigan State University for their help in installing the system.
The technical computer personnel at Afeka were a great help. The participation in this conference
was financially supported by Afeka Engineering College. Comments and suggestions
by my wife Hana Goldman helped considerably in improving the paper.

\bibliographystyle{IEEE}

\end {document}